\documentstyle[psfig,preprint,tighten,eqsecnum,floats,aps]{revtex}

\def\spi{i^o}
\def\ut#1{\rlap{\lower1ex\hbox{$\sim$}}{#1}}
\def\l{\ell}
\def\N{{\cal N}}

\def\bF{{\bf F}}

\def\bA{{\bf A}}
\def\M{{\bf M}}

\def\Q{{\tilde{Q}}}
\def\pb#1{\rlap{\lower1.5ex\hbox{$\longleftarrow$}}{#1}}
\def\dpb#1{\rlap{\lower1.5ex\hbox{$\Longleftarrow$}}{#1}}
\def\spb#1{\rlap{\lower1.0ex\hbox{$\leftarrow$}}{#1}}
\def\sdpb#1{\rlap{\lower1.0ex\hbox{$\Leftarrow$}}{#1}}
\def\d{{\rm d}}

\def\bar{\overline}
\def\ba{\begin{eqnarray}}
\def\ea{\end{eqnarray}}
\def\be{\begin{equation}}
\def\ee{\end{equation}}
\def\={\mathrel{\widehat\mathalpha{=}}}
\def\puto#1{\rlap{\raise.5ex\hbox{\char'27}}{#1}}

\preprint{\vbox{\baselineskip=12pt
\rightline{gr-qc/9910068}
\rightline{CGPG-99/10-2}}}

\begin{document}
\draft
\title{Laws Governing Isolated Horizons:\\
Inclusion of Dilaton Couplings}
\author {Abhay\ Ashtekar\thanks{E-mail:
 ashtekar@gravity.phys.psu.edu}${}^{1,3}$ and 
Alejandro\ Corichi\thanks{E-mail: 
corichi@nuclecu.unam.mx}${}^{1,2,3}$
}
\address{1. Center for Gravitational Physics and Geometry \\
Department of Physics, The Pennsylvania State University \\
University Park, PA 16802, USA}

\address{2. Instituto de Ciencias Nucleares\\
Universidad Nacional Aut\'onoma de M\'exico\\
A. Postal 70-543, M\'exico D.F. 04510, M\'exico.}

\address{3. Institute for Theoretical Physics,\\ 
University of California, Santa Barbara, CA 93106, USA}
\maketitle

\begin{abstract}
Mechanics of non-rotating black holes was recently generalized by
replacing the static event horizons used in standard treatments with
`isolated horizons.'  This framework is extended to incorporate
dilaton couplings.  Since there can be gravitational and matter
radiation outside isolated horizons, now the fundamental parameters of
the horizon, used in mechanics, must be defined using only the
\textit{local} structure of the horizon, without reference to
infinity.  This task is accomplished and the zeroth and first laws are
established. To complement the previous work, the entire discussion is
formulated tensorially, without any reference to spinors.
\end{abstract}
\pacs{Pacs: 04070B, 0420}

\section{Introduction}
\label{sec1}

The zeroth and first laws of black hole mechanics refer to
equilibrium situations and small departures therefrom.  Therefore, in
the standard treatments \cite{1,2,3,4,mh} one restricts oneself to
stationary space-times admitting event horizons and perturbations off
such space-times.  While this simple idealization is a natural
starting point, from physical considerations it seems overly
restrictive.  (See \cite{ack,abf1} and especially \cite{abf2} for a
detailed discussion.).  A framework, which is tailored to more
realistic physical situations was introduced in \cite{ack} and the
zeroth and first laws were correspondingly extended in
\cite{abf1,abf2}.  This framework generalizes black hole mechanics in
two directions.  First, the notion of event horizons is replaced by
that of `isolated horizons'; while the former can only be defined
retroactively, after having access to the entire space-time history,
the latter can be defined quasi-locally.  Second, the underlying
space-time need not admit any Killing field; isolated horizons need
not be Killing horizons.  The static event horizons normally used in
black hole mechanics \cite{1,2,3} and the cosmological horizons in de
Sitter space-times \cite{5} are all special cases of isolated
horizons.  However, because we can now admit gravitational and matter
radiation, there are many more examples.  In particular, while the
space ${\cal S}$ of static space-times admitting event horizons in the
Einstein-Maxwell theory is finite dimensional, the space ${\cal IH}$
of space-times admitting isolated horizons is \textit{infinite}
dimensional \cite{abf2}.

In the discussion of zeroth and first laws in \cite{abf2}, two
restrictive assumptions were made.  First, the horizon was assumed to
be non-rotating.  Second, while a rather general class of matter
fields was allowed, it was assumed that the only relevant charges
---i.e., hair--- are the standard electric and magnetic ones.  In this
note, we will continue to work with non-rotating horizons but weaken
the second assumption by allowing a dilaton field and a general
coupling, parameterized as usual by a (non-negative) coupling constant
$\alpha$.  By setting $\alpha =0$, one recovers the standard
Einstein-Maxwell-Klein-Gordon system (which was included in
\cite{abf2}). For $\alpha =1$ (in units in which the string tension
is set equal to one) one obtains the action widely used in the study
of the low energy limit of string theory. For all values of $\alpha$,
we will be able to extend the isolated horizon framework and prove the
zeroth and first laws.

The main challenge encountered in this extension is the following.  In
the standard treatment, static black holes in dilaton gravity are
characterized by three parameters: the mass $M$, the (modified)
electric charge $\Q$ (or, the magnetic charge $P$) and a dilaton
charge $\phi_\infty$.  Of these, $M$ and $\phi_\infty$ are defined at
\textit{infinity} (while $\Q$ and $P$, being absolutely conserved, can
be defined on any 2-sphere).  In the more general context of isolated
horizons, gravitational radiation and matter fields may be present
between the horizon and infinity.  Now $M$ and $\phi_\infty$ carry
information about these fields and can no longer be regarded as
intrinsic parameters of the horizon itself.  Therefore, the laws of
mechanics can not refer to them; we need new intrinsic parameters.  In
the Einstein-Maxwell case, this was accomplished \cite{abf1,abf2} by
replacing $M$ with the area $a_\Delta$ of the isolated horizon
$\Delta$ (in this case, $\phi_\infty=0$).  A natural strategy now is
to continue to use $a_\Delta$ in place of $M$ and use the value
$\phi_\Delta$ of the dilaton field \textit{on} the horizon in place of
$\phi_\infty$.  However, it is far from obvious that this strategy is
viable.  For, there are at least four conditions to meet.  First, the
isolated horizon boundary conditions should tell us that the field
$\phi_\Delta$ is constant on $\Delta$.  Second, we should be able to
define surface gravity $\kappa$ in absence of a Killing field and show
that it is constant on \textit{any} isolated horizon, in spite of the
fact that the values of fields such as the Weyl component $\Psi_4$ and
the Maxwell component $\phi_2$ can be `dynamical' on isolated
horizons.  Third, to have a meaningful formulation of the first law,
the Hamiltonian strategy of \cite{abf2} for defining the mass
$M_\Delta$ of the isolated horizon should go through and enable us to
express $M_\Delta$ in terms only of the horizon parameters, in spite
of the fact that values of fields such as $\Psi_4$ and $\Phi_2$ on
$\Delta$ are not determined by the horizon parameters.  Finally, the
functional dependence of the horizon mass $M_\Delta$, the surface
gravity $\kappa$ and the electric potential $\Phi$ on the horizon
parameters $a_\Delta, \Q_\Delta$ (or, $P_\Delta$) and $\phi_\Delta$
should be such that the first law holds.  We will show that all these
conditions can be met.

The organization of the paper is as follows.  For convenience of
readers who may not be familiar with dilaton couplings, in Section
\ref{s2} we briefly recall the relevant results from literature.  In
Section \ref{s3}, we specify the boundary conditions defining
non-rotating isolated horizons in dilaton gravity and summarize
their consequences which are needed for the main results.  Section
\ref{s4} introduces surface gravity $\kappa$ and establishes the
zeroth law while Section \ref{s5} introduces the mass $M_\Delta$ and
establishes the first law.  Since the underlying strategy in sections
\ref{s4}and \ref{s5} is the same as in \cite{abf2}, we will use the
same notation (which, incidentally, differs in some minor ways from
the conventions used in \cite{ack}) and refrain from repeating the
detailed arguments and proofs given in \cite{abf2}.  Rather, we will
recall the main conceptual steps and emphasize the new issues and
subtleties that arise due to dilaton couplings. Also, we will use
this opportunity to make some new observations which hold also in
cases considered in \cite{abf2}. Finally, while the earlier treatments
were spinorial, in this paper, we will complement that discussion by
working only with tensors.

\section{Dilaton coupling and static black holes}
\label{s2}

\subsection{Einstein-Maxwell-Dilaton System}
\label{s2.1}

In dilaton gravity, (in the so-called Einstein frame) the gravitational 
part of the action, $S_{\rm Grav}$ is the standard one \cite{ack,abf2}.  
The matter part of the action may be less familiar.  It is given by:
\be S_{\rm Dil} (\phi, A) =-\frac{1}{16\pi}\int_{{\cal M}}\sqrt{-g} 
[2(\nabla\phi)^2 + e^{-2\alpha\phi}\bF_{ab}\bF^{ab}]\d^4x\label{dil:act} 
\ee
where $\alpha $ is a free parameter which governs the strength of the 
coupling of the dilaton field $\phi$ to the Maxwell field $\bF_{ab}$.  
Since the action is invariant under simultaneous changes of sign of $\phi$ 
and $\alpha$, without loss of generality, one can restrict oneself to 
non-negative values of $\alpha$.  If $\alpha=0$, we recover the 
Einstein-Maxwell-Klein-Gordon system. The total action 
$S_{\rm Tot}$ is given by
\be S_{\rm Tot}=S_{\rm Grav} + S_{\rm Dil}. \ee  
The equations of motion that follow from 
$S_{\rm Tot}$ are:
\ba 
\nabla_a {}^\star \bF^{ab} = 0,\quad\quad
\nabla_a(e^{-2\alpha\phi}\bF^{ab})=0,\label{dil:eom1}\\
\nabla^2\phi+\frac{\alpha}{2}e^{-2\alpha\phi}\bF^2=0,\quad {\rm and}
\label{dil:eom2}\\
R_{ab}= G\left[2\nabla_a\phi\nabla_b\phi+2e^{-2\alpha\phi}
\bF_{ac}{\bF_b}^c -\frac{1}{2}
g_{ab}e^{-2\alpha\phi}\bF^2  \right],
\label{dil:eom3}
\ea
where $\bF^2=\bF_{ab}\bF^{ab}$. 

Since $\d \bF =0$ by definition of $\bF$, the magnetic charge,
\be
P[{S}] := \frac{1}{4\pi}\, \oint_S \bF
\ee
is absolutely conserved, i.e., is independent of the particular choice
of the 2-sphere $S$, in a given homology class, made in its
evaluation.  Note however that $\d\, {}^\star \bF \not= 0$, whence the
standard electric charge 
\be 
Q_{\rm std} := \frac{1}{4\pi}\,\oint_S {}^\star\bF
\ee
(with ${}^\star \bF_{ab} = \frac{1}{2}\, \epsilon_{abcd}\bF^{cd}$) is
\textit{not} absolutely conserved. The effective current $J^a$
depends on the Maxwell field itself and is given by $J^a = 2\alpha \,
[\exp{-2\alpha\phi}]\, \nabla_b \phi\, F^{ab}$.  Hence, $Q_{\rm std}$
satisfies a balance equation:
\be Q_{\rm std}[S_1] - Q_{\rm std}[{S_2}]= \frac{\alpha}{2\pi}
\int_{M_{12}} \d\phi\wedge{}^\star \bF
\ee
where $M_{12}$ is the 3-surface with boundaries $S_1$ and $S_2$.
However, it is
obvious from (\ref{dil:eom1}) that the 2-form $(\exp\,
{-2\alpha\phi})\,\, {}^\star\bF $ is closed.  Therefore, there is in
fact a conserved electric charge,
\be
{Q}_{\rm cons}[S]:=\frac{1}{4\pi}\oint_{S} e^{-2\alpha\phi}\,\,\,
{}^\star\bF
\, .\ee
Again, $Q_{\rm cons}[S]$ depends only on the homology class of $S$.
Finally, in the asymptotically flat context of primary interest to
this paper, the requirement that the 4-momentum be finite leads one to
a natural boundary condition on the dilaton field at spatial infinity:
$\phi$ must approach a constant, $\phi_\infty$.  This constant
provides a `dilatonic charge' $\phi_\infty$ associated with the
solution.
\footnote{In the literature, dilatonic charge $D$ is generally
defined only in stationary space-times. The new information in $D$,
not already contained in the mass $M$ and charge $Q_{\rm cons}$, is
$\phi_\infty$.}
Note that the values of $Q_{\rm
std}[S_\infty]$ ,$Q_{\rm cons}[S_{\infty}]$ and $\phi_\infty$ are
related via
\be Q_{\rm cons}[S_{\infty}] = e^{-2\alpha\phi_\infty} \,\,
Q_{\rm std}[S_\infty]
\ee
so that, in the parameterization of the solution, one can replace  
$\phi_\infty$ with $Q_{\rm std}[S_\infty]$.

Of special interest are two symmetries of the theory. The first is a 
natural generalization of the standard duality rotation of the 
Maxwell theory:
\be \label{duality} 
(\phi, \bF_{ab}, g_{ab}) \,\, \mapsto \,\, {\cal D}\, (\phi, \bF_{ab},
g_{ab}) = (-\phi, {}^\star\bF_{ab}, g_{ab}) \, .
\ee
As in the standard Maxwell theory, the action fails to be invariant
under this transformation since $(\exp\, -2\alpha\phi)\,\, \bF^2 \,
\mapsto - (\exp -2\alpha\phi)\,\, \bF^2$.  However, it is obvious
from the equations of motion that if $(\phi, \bF, g)$ is a solution,
so is the duality rotated triplet ${\cal D}\,(\phi, \bF, g)$.  This
fact is generally exploited to restrict oneself to the sector of the
theory in which the magnetic charge vanishes: Properties of solutions
with a magnetic charge can be obtained simply by a duality rotation of
the solution with a corresponding electric charge.%
\footnote{Note however that ${\cal D}$ is a \textit{discrete}
transformation and maps solutions with pure electric charge to those
with pure magnetic charge.  The analog of the continuous
transformations $(\bF,g)\, \mapsto (\cos\theta \bF+
\sin\theta\,{}^\star\bF,\, g)$ parameterized by an angle $\theta$ of
the Einstein-Maxwell theory does not seem to exist in dilaton gravity.
Furthermore, solutions with both electric and magnetic charge are
known \cite{kall} only for $\alpha=1$ and $\alpha=\sqrt{3}$, and they
were {\it not} constructed using a symmetry of the theory .}

The second symmetry is continuous and corresponds to a constant shift in 
$\phi$:
\be \label{trans} 
(\phi, \bF_{ab}, g_{ab}) \mapsto {\cal K}\,(\phi, \bF_{ab}, g_{ab}) =
(\phi + k,\, \exp k\alpha \, \bF_{ab}, g_{ab})
\ee
where $k$ is a real constant.  It is easy to see that this
transformation leaves the action invariant and hence also maps
solutions to solutions.  This symmetry is often exploited to restrict
oneself to the sector of the theory in which $\phi_\infty =0$.  Again,
properties of solutions with non-zero $\phi_\infty$ can be obtained by
applying the appropriate map ${\cal K}$.

{}From now on, for notational simplicity, we set 
$Q_{\rm std} [S] = {Q}[S]$ and $Q_{\rm cons} = {\Q}$

\subsection{Static black holes}
\label{s2.2}

Static, spherically symmetric solutions to dilaton gravity are special
cases of a general class of black holes first discovered by Gibbons
\cite{gg} and Gibbons and Maeda \cite{gm}.  (These solutions were 
independently discovered by Garfinkle, Horowitz and Strominger
\cite{ghs}.) In the case $\alpha =1$,  they are the unique static
black hole solutions of dilaton gravity \cite{m}. For general
$\alpha$, we will refer to them as the Gibbons-Maeda (GM) solutions.

By making an appeal to the duality rotation ${\cal D}$ and the
constant shift transformation ${\cal K}$, one restricts oneself to the
case in which the magnetic charge $P$ and the dilaton charge
$\phi_\infty$ vanish.  Then, the general solution is conveniently
parameterized by two numbers, $r_{\pm}$ (with $\infty > r_+\ge r_-
>0$), related to the ADM mass $M$ and conserved electric charge $\Q$
via:
\ba
GM&=&\frac{r_+}{2}+\left[\frac{1-\alpha^2}{1+\alpha^2}\right]
\frac{r_-}{2}\\
G{\Q^2} &=&\left[\frac{r_-\,r_+}{1+\alpha^2}\right]
\ea
The solution is given by:
\ba
\d s^2&=&-\lambda^2\d t^2+\lambda^{-2}\d r^2+R^2\d \Omega\\
{\bf F}&=&\frac{\Q}{r^2}\;\d t\wedge\d r, \quad \, {\rm and} \\
e^{2\alpha\phi}&=&\left[1-\frac{r_-}{r}\right]^{2\alpha^2/(1+\alpha^2)}\, .
\ea
Here, $r$ is a radial coordinate, related to the geometrical radius
$R$ of 2-spheres spanned by the orbits of the rotational Killing
fields via
\be
R=r\left[1-\frac{r_-}{r}\right]^{\alpha^2/(1+\alpha^2)}
\ee
and $\lambda$ is a function of $r$, given by
\be
\lambda^2=\left[1-\frac{r_+}{r}\right]\left[1-\frac{r_-}{r}
\right]^{(1-\alpha^2)/(1+\alpha^2)}.
\ee
When $\alpha=0$, the dilaton field $\phi$ vanishes and the solution
reduces to the Reissner-Nordstr\"om solution of the Einstein-Maxwell
theory.  The case $\alpha=1$ is of interest to string theory.  For all
$\alpha$, $r=r_+$ is an event horizon and for all \textit{non-zero}
values of $\alpha$, there is a curvature singularity at $r= r_-$.  For
$\alpha=0$, $r=r_-$ is a non-singular inner horizon.  Finally, the
extremal black holes correspond to $r_+ = r_-$ (or
$Q^2=G(1+\alpha^2)\,\,M^2$).  The horizon of these extremal black
holes is singular except when $\alpha=0$.
 
In this 2-parameter family of black holes, one can directly compute
the acceleration of the Killing field $t^a$ (with $t^a\partial_a =
\partial/\partial t$) on the horizon to obtain an expression for the
surface gravity $\kappa$. We have:
\be \label{kappa1} 
\kappa =\frac{1}{2r_+}\left[1-\frac{r_-}{r_+}
\right]^{(1-\alpha^2)/(1+\alpha^2)}\, .
\ee
{}From (\ref{kappa1}) it is obvious that the zeroth law holds.
Furthermore, $\kappa>0$ for all non-extremal black holes.  In the
extremal case, since the event horizon is singular unless $\alpha=0$,
surface gravity is defined unambiguously only if $\alpha =0$ (and is
zero in this case).  However, if $\alpha\le 1$, one can define
$\kappa$ by a limiting procedure and show that $\kappa =0$ for
$\alpha\le 1$ and $\kappa = \frac{1}{4M}$ if $\alpha=1$.

Next, let us consider the first law.  It is straightforward to check
that, for any variation $\delta$ within this family of solutions, we
have
\be \label{1law1}
\delta{M}\, = \,\frac{1}{8\pi G} \kappa\, \delta{a_{\rm hor}} + 
\Phi \delta {\Q}\, , \ee
where $a_{\rm hor}$ is the area of the horizon and $\Phi :\= \bA_a t^a
\equiv {\Q}/r_+$ is the value of the natural scalar potential at the
horizon.

To conclude, let us return to the symmetry transformations.  Since the
ADM mass $M$ and the surface gravity $\kappa$ are determined entirely
by the metric and the metric does not change under ${\cal D}$ or
${\cal K}$, neither do $M$ and $\kappa$.  On the other hand, the
electric and magnetic charges do change.  The effect of ${\cal D}$ is
the same as in Maxwell's theory.  Let us therefore focus on ${\cal
K}$.  Let us begin with a Gibbons-Maeda solution $(\phi^o, \bF^o_{ab},
g^o_{ab})$ and set ${\cal K}\,(\phi^o, \bF^o_{ab}, g^o_{ab}) = (\phi,
\bF_{ab}, g_{ab})$.  The resulting solution does not belong to the
Gibbons-Maeda family since $\phi \mapsto k \not= 0$ at infinity.  The
parameters of the new solution are $(r_\pm = r_\pm^o,\,
\phi_\infty\!=k)$.  Thus, the (trivially) generalized Gibbons-Maeda
(gGM) solutions are characterized by \textit{three} parameters,
$r_\pm, k$ with $0<r_- \le r_+ <\infty$ and $-\infty <k <\infty$,
rather than just two.  In terms of these, the physical quantities
associated with the black-hole are given by:
\be 
M= M^o, \,\, a_{\rm hor} = a_{\rm hor}^o, \,\,
\kappa = \kappa^o, \,\, \Phi = \Phi^o\, e^{\alpha k}\,\,\, {\rm and} 
\,\,\, {\Q} = {\Q}^o\, e^{-\alpha k}.
\ee
Consequently, for this family, the first law becomes:
\ba \label{gGM1law} 
\delta{M} &=& \frac{1}{8\pi G}\, \kappa \,\delta{a_{\rm hor}} + 
\Phi \delta{\Q} + \alpha \Phi \Q \, \delta k \nonumber\\ 
&=&\frac{1}{8\pi G}\, \kappa \delta{a_{\rm hor}} +
\hat\Phi \delta{\hat{Q}} \ea
where $\hat{Q} = {\Q}e^{\alpha k}$ (i.e. $\hat{Q}^2=\Q\,Q[S_\infty]$) and
$\hat{\Phi} = \Phi e^{-\alpha k}$.  We will return to these
generalized Gibbons-Maeda solutions in Sections \ref{s4} and \ref{s5}.

\section{Boundary conditions and their consequences}
\label{s3}

We are now ready to consider situations which are not necessarily
static and introduce the notion of isolated horizons $\Delta$ for
dilaton gravity.  The basic boundary conditions defining $\Delta$
are the same as those introduced in \cite{ack,abf2}.  However, we will
use this opportunity to present them \textit{without} recourse to
spinors.  While this formulation is less convenient for detailed
calculations such as the ones performed in \cite{abf2}, it makes
the meaning of the underlying assumptions more transparent.

Let us begin by introducing some notation. Fix any null surface
${\N}$, topologically $S^2\times R$, and consider foliations of $\N$
by families of 2-spheres transversal to its null normal.  Given a
foliation, we parameterize its leaves by $v={\rm const}$ such that $v$
increases to the future and set $n = -\d v$.  Under a
reparametrization $v\mapsto F(v)$, we have $n\mapsto F'(v) n$ with
$F'(v) >0$.  Thus, every foliation comes equipped with an equivalence
class $[n]$ of normals $n$ related by rescalings which are constant on
each leaf.%
\footnote{These 1-form fields $n_a$ are defined intrinsically on $\N$.
We can extend each $n_a$ to the full space-time uniquely by demanding
that the extended 1-form be null. However, in this paper, we will not
need this extension.}
Now, given any one $n$, we can uniquely select a vector field $\l$
which is normal to $\N$ and satisfies $\l^a n_a = -1$. (Thus, $\l$ is
future-pointing.  If we change the parameterization, $\l^a$ transforms
via: $\l^a \mapsto (F'(v))^{-1} \l^a$.  Thus, given a foliation, we
acquire an equivalence class $[\l^a, n_a]$ of pairs, $(\l^a, n_a)$ ,
of vector fields and 1-forms on $\N$ subject to the relation $(\l^a,
n_a) \sim (G^{-1} \l, G n)$, where $G$ is any positive function on
$\N$ which is constant on each leaf of the foliation.  Given a pair
$(\l, n)$ in the equivalence class, we introduce a complex vector
field $m$ on $\N$, tangential to each leaf in the foliation, such that
$m\cdot \bar{m} = 1$.  (By construction, $m\cdot \l = m\cdot n = 0$ on
$\N$.)  The vector field $m$ is unique up to a phase factor.  With
this structure at hand, we now introduce the main Definition.

\bigskip\noindent {\it Definition:} The internal boundary $\Delta$ of a 
space-time $(\M, g_{ab})$ will be said to represent {\it a
non-rotating isolated horizon} provided the following conditions hold%
\footnote{Throughout this paper, the symbol
{${\mathrel{\widehat\mathalpha{=}}}$} will denote equality at points
of $\Delta$.  For fields defined throughout space-time, an under-arrow
will denote pull-back to $\Delta$. The part of the Newman-Penrose
framework \cite{pr} used in this paper is summarized in the Appendices
A and B of \cite{abf2}.}:
\begin{itemize} 
\item{(i)} {\it Manifold conditions:} $\Delta$ is a null surface,
topologically $S^{2}\times R$.
\item{(ii)} {\it Dynamical conditions:} All field equations 
hold at $\Delta$.
\item{(iii)} {\it Main conditions:} $\Delta$ admits a foliation such that 
the Newman-Penrose coefficients associated with the corresponding direction 
fields $[\l,n]$ on $\Delta$ are subject to the following conditions:\\
(iii.a) $\rho \= -\bar{m}^a {m}^b \nabla_a\l_b$, the expansion of $[\l]$, 
vanishes on $\Delta$.\\
(iii.b) $\lambda \= \bar{m}^a \bar{m}^b \nabla_a n_b$ and $\pi \= \l^a
\bar{m}^b \nabla_a n_b$ vanish on $\Delta$ and the expansion $\mu :=
m^a\bar{m}^b \nabla_a n_b$ of $n$ is negative
\footnote{For simplicity, in this paper we focus on black-hole-type 
horizons rather than cosmological ones. To incorporate interesting
cosmological horizons, one has to weaken this condition and allow 
the possibility that $\mu$ is everywhere positive on $\Delta$. See
\cite{abf2}.}
and constant on each leaf of the foliation.
\item{(iv)} {\it Conditions on matter:} The Maxwell field $\bF$ is such 
that 
\be \phi_1 \= \frac{1}{2}\, m^a\bar{m}^b (\bF - i\, {}^\star\!\bF)_{ab}
\ee
is constant on each leaf of the foliation introduced in condition (iii).
\end{itemize}

The first two conditions are quite tame: (i) simply asks that $\Delta$
be null and have appropriate topology while (ii) is completely
analogous to the dynamical condition imposed at infinity.  As the
terminology suggests, (iii.a) and (iii.b) are the most important
conditions.  Note first that, if a pair $(\l,n)$ in the equivalence
class $[\l,n]$ associated with the foliation satisfies these
conditions, so does any other pair, $((G(v))^{-1} \l ,\, G(V) n)$.
Thus, the conditions are well-defined.  They are motivated by the
following considerations.  Condition (iii.a) captures the idea that
the horizon is isolated without having to refer to a Killing field.
In particular, it implies that the area of each 2-sphere leaf in the
foliation be the same.  We will denote this area by $a_\Delta$ and
define the \textit{horizon radius} $R_\Delta$ via $a_\Delta = 4\pi
R^2_{\Delta}$.  (In previous papers \cite{ack,abf1,abf2}, the horizon
radius was denoted by $r_\Delta$.  We have changed the notation to
avoid possible confusion with the (non-geometric) radial coordinate
$r$ generally used in the discussion of dilaton black holes.)

Condition (iii.b) has three sets of implications.  First, one can show
that if, as required, one can find a foliation of $\Delta$ satisfying
(iii.b), that foliation is \textit{unique}.  (In the gGM family, as
one might expect, this condition selects the foliation to which the
three rotational Killing fields are tangential.)  Second, it implies
that the imaginary part of (the Newman-Penrose Weyl component)
$\Psi_2$, which captures angular momentum, vanishes and thus restricts
us to {\it non-rotating} horizons. Third, the requirement that the
expansion $\mu$ of $n^a$ be negative implies that $\Delta$ is a
\textit{future} horizon rather than past \cite{sh}. Finally, consider the
spherical symmetry requirement on the Maxwell field component
$\phi_1$.  While this condition is a strong restriction, it can be
motivated as follows.  Conditions (i) -- (iii) imply that $\phi_0 \=
-\l^am^b\bF_{ab}$, the `radiative part of the Maxwell field traversing
$\Delta$' vanishes (confirming the interpretation that $\Delta$ is
isolated). If the `radial derivative' of $\phi_0$ also vanishes at
$\Delta$ ---i.e., heuristically, if there is no flux of
electro-magnetic radiation across $\Delta$ also to the next order---
condition (iv) is automatically satisfied.  (For further motivation
and remarks on these conditions, see \cite{ack,abf2}.)  Note that
there is no explicit restriction on the dilaton field at $\Delta$.

Since these conditions are \textit{local} to $\Delta$, as indicated in
the Introduction, the notion of an isolated horizon is quasi-local; in
particular, one does not need an entire space-time history to locate
an isolated horizon.  Furthermore, the boundary conditions allow for
presence of radiation in the exterior region.  Therefore, space-times
admitting isolated horizons need not admit any Killing field
\cite{jl}.  Indeed, the space ${\cal I}$ of solutions to field
equations admitting isolated horizons is infinite dimensional.
(Strategies for construction of such solutions are discussed in
\cite{abf2}).

In spite of this generality, boundary conditions place surprisingly
strong restrictions on the structure of various fields \textit{at}
$\Delta$.  Let us begin with conditions on the dilaton and Maxwell
fields.  The stress-energy tensor $T_{ab}$ of $(\phi, \bF)$ satisfies
the dominant energy condition.  Hence, on $\Delta$, $-T_{ab}\l^b$ is a
future directed, causal vector field.  Now, using the Raychaudhuri
equation and field equations \textit{at} $\Delta$ (condition (ii) of
the Definition), we conclude $T_{ab}\l^a\l^b \=0$.  By expanding out
this expression (see Eq (\ref{dil:eom3})) we obtain
\be \label{3.2}
\dot\phi \equiv {\cal L}_\l\, \phi \=0  \quad{\rm and}\quad
\bF \= \phi_1 [\l\wedge n - m\wedge \bar{m}] + \phi_2 [m\wedge \l]
+{\rm CC} 
\ee
for \textit{some} complex functions $\phi_1$ and $\phi_2$ (the
Newman-Penrose components of $\bF$) on $\Delta$, where ${\rm CC}$
stands for `the complex conjugate term'.  These equations say that
that there is no flux of dilatonic or electro-magnetic radiation
across $\Delta$. Next, since $-T_{ab}\l^a$ is a future pointing
causal vector and $T_{ab}\l^a\l^b\=0$, it follows that $T_{ab}{\l}^a
m^b\=0$. Finally, the main conditions (iii) in the Definition imply
that $R_{ab}m^am^b \=0$ \cite{ack,abf2} and, since the field equations
hold at $\Delta$, we conclude $T_{ab}m^am^b \= 0$. 
This implies that the field $\phi$ is constant on $\Delta$.
We will denote this constant by $\phi_\Delta$ and refer to it as
\textit{the dilatonic charge of the isolated horizon}.  Finally,
condition (iv) in the Definition implies 
\be \phi_1 \= e^{2\alpha \phi_\Delta}\,\, \frac{2\pi}{a_\Delta}\,
{\Q}_\Delta \, ,\ee
where $\phi_\Delta$ is the value of the dilaton field on $\Delta$ and
$\Q_\Delta \equiv \Q$ is the conserved electric charge.  Thus the
boundary conditions severely restrict the form of matter fields at
$\Delta$.  The dilaton field $\phi$ is constant on $\Delta$, the
component $\phi_0 = -\l^a m^b \bF_{ab}$ of the Maxwell field vanishes
and the component $\phi_1$ is completely determined by the dilaton and
(conserved) electric charge.  However, the component $\phi_2$ of the
electro-magnetic field is unconstrained.

Restrictions imposed on space-time curvature at $\Delta$ are
essentially the same as in Ref \cite{abf2}.%
\footnote{This is because these restrictions were obtained assuming 
rather general conditions on the matter stress-energy which are satisfied 
in dilaton gravity.  The derivation of some of these results involve long 
calculations and a topological result on the Chern-class of the $SO(2)$ 
connection associated with the dyad $(m,\bar{m})$. See \cite{abf2}.}
Results relevant to this paper can be summarized as follows.  In the
Newman-Penrose notation, for the Ricci tensor components, we have:
\ba \label{ricci}
\Phi_{00} &=& \frac{1}{2}\, R_{ab} \l^a\l^b \= 0, \quad
\Phi_{01} = \frac{1}{2}\, R_{ab}\l^a m^b \= 0,
\Phi_{02} = \frac{1}{2}\, R_{ab} m^a m^b \= 0,
\nonumber\\
\Phi_{11} &=& \frac{1}{4}R_{ab} (\l^an^b +m^a\bar{m}^b) \=
- 8\pi^2G \, e^{2\alpha\phi_\Delta}\, \left( 
\frac{\Q_\Delta}{a_\Delta}\right)^2, \quad R \= 0 ,
\ea
where $R$ is the scalar curvature. The Weyl tensor components satisfy
\ba\label{weyl}
\Psi_0 &=& C_{abed}\l^am^b\l^cm^d \= 0, \qquad
\Psi_1 = C_{abed}\l^am^b\l^cn^d \= 0\, \nonumber\\
\Psi_2 &=& C_{abcd}\l^a m^b\bar{m}^c n^d \= \Phi_{11}  - 
\frac{2\pi}{a_\Delta} 
\ea
Furthermore, 
\be
\Psi_3 \= \Phi_{21}, \quad{\hbox {\rm that is}}\quad
C_{abcd}\l^an^b\bar{m}^cn^d \= \frac{1}{2} R_{ab}\bar{m}^an^b.
\ee
This structure is the same as in the gGM solutions discussed at the
end of Section \ref{s2}.  However, on a general isolated horizon,
other curvature components may be `dynamical', i.e., vary along the
integral curves of $\l$.

In view of this structure of various fields on $\Delta$, it
is natural to use the triplet $R_\Delta, \tilde{Q}_\Delta,
\phi_\Delta$ ---the horizon radius, the conserved charge and the value
of the dilaton field on the horizon--- to parameterize general
isolated horizons. In the gGM solutions, these parameters are subject
to the restriction $R_\Delta^2 \ge (1+\alpha^2)G \Q^2 \exp
(2\alpha\phi_\Delta)$.  In the more general context of isolated
horizons, we will restrict the range of parameters by the same
condition. Unlike the ADM mass $M$, the total
electric charge $Q_\infty$ and the dilatonic charge $\phi_\infty$,
these parameters are \textit{local} to $\Delta$.  In the special case
of gGM solutions, this triplet is uniquely determined by the
parameters $r_\pm, k\equiv \phi_\infty$ used in Section \ref{s3} and,
reciprocally, determine $r_\pm, k$ uniquely.

We will conclude this section with a remark. There exist in the
literature other families of static, spherically symmetric solutions
with Maxwell and dilaton fields with Killing horizons which, however,
fail to be asymptotically flat (or anti-de Sitter) \cite{chm}. In the
general case, the field equations they obey contain a potential term
$V(\phi)$ for the dilaton field and reduce to those given in Section
\ref{s2.1} only in the case when $V(\phi) = 0$. However, for all
potentials, our boundary conditions are satisfied at the Killing
horizons of these solutions; thus those horizons are, in particular,
isolated horizons. Furthermore, for \textit{general} isolated horizons
in any of these theories, even without Killing fields, all conclusions
of this section continue to hold (even though, for certain potentials
$V(\phi)$, the stress-energy tensor does not satisfy the dominant
energy condition).  Isolated horizons offer a natural home for such
black holes since, unlike event horizons, they do not refer to the
structure at infinity.

\section{Surface gravity and the zeroth law}
\label{s4}

In each gGM solution there is a unique time-translational Killing
field $t^a$ which is unit at infinity.  As usual, surface gravity
$\kappa_{\rm GM}$ is defined in terms of its acceleration at the
horizon: $t^a \nabla_a t^b\= \kappa_{\rm GM}\, t^b$.  In terms of the
gGM parameters $(r_{\pm}, k)$, $\kappa_{\rm GM}$ is given by
(\ref{kappa1}). {}From the perspective of the isolated horizon
framework, $\kappa$ is the acceleration of the properly normalized
null normal $\l^a$ to $\Delta$ \cite{abf1,abf2}. In the gGM solutions,
$\Delta$ happens to be a Killing horizon and we can select a unique
vector field $\l^a$ from the the equivalence class $[\l^a]$ simply by
setting $\l^a \= t^a$. Then $\kappa_{\rm GM}$ is the acceleration of
this specific $\l^a$.  In the case of general isolated horizons, the
challenge is to find a prescription to single out a preferred $\l^a$,
without reference to any Killing field.  The strategy we adopt is
identical to that used for isolated horizons without a dilaton
coupling in \cite{abf1,abf2}.  However, unlike in those references, we
will now proceed in two steps to bring out a general conceptual issue.

In the first step, we will normalize $\l$ {\it only} up to a constant,
leaving a rescaling freedom $\l \mapsto \l^\prime = c\l$, where $c$ is
a constant on $\Delta$ but may depend on the parameters $R_\Delta,
{\Q}_\Delta, \phi_\Delta$ of the isolated horizon.  For each such
$\l$, we can define the surface gravity $\kappa_{\l}$ {\it relative to
that} $\l$ via $\l^a\nabla_a \l^b \= \kappa_{\l} \l^b$.  Rescaling of
$\l$ now induces to a `gauge transformation' in $\kappa$: $\kappa_{\l}
\mapsto \kappa_{\l^\prime} = c \kappa_{\l}$.  (Recall that in the
general Newman-Penrose framework, $\kappa$ is a connection component
and therefore undergoes the standard gauge transformations under a
change of the null tetrad.  By fixing $\l$ up to a constant rescaling,
we have reduced the general gauge freedom to that of a constant
rescaling.)  Since the zeroth law only says that the surface gravity
is constant on $\Delta$, if it holds for one $\l$, it holds for every
$\l^\prime = c\l$.  Thus, for the zeroth law, it is in fact
\textit{not} essential to get rid of the rescaling freedom.

Recall that the isolated horizon is naturally equipped with
equivalence classes $[\l,n]$ of vector and co-vector fields, subject
to the relation: $(\l, n) \sim (G^{-1}\l, G n)$ for any positive
function $G \equiv G(v)$ on $\Delta$.  Our first task is to reduce the
freedom in the choice of $G(v)$ to that of a constant.  We use the
same strategy as in \cite{abf1,abf2}.  (For motivation, see
\cite{abf2}.)  Recall that $\mu$, the expansion of $n$ is strictly
negative  and constant on each leaf of the preferred
foliation; $\mu \equiv \mu(v) <0$.  It is easy to verify that
\be n^a \mapsto G(v) n^a \,\,{\rm implies}\quad \mu \mapsto G(v)
\mu(v).  \ee
Hence, we can \textit{always} use the $G(v)$ freedom to set $\mu \=
{\rm const}$.  This condition restricts the family of $(\l,n)$ pairs
and reduces the equivalence relation to $(\l,n) \sim (c\l, c^{-1}n)$
where $c$ is any constant on $\Delta$. We will denote the restricted
equivalence class by $[\l,n]_R$. In the second step, we will
\textit{fix} the numerical value of $\mu$ in terms of the parameters
of the isolated horizon and eliminate the rescaling freedom
altogether, thereby selecting a canonical pair $(\l, n)$ on each
isolated horizon.

With the equivalence class $[\l,n]_R$ at our disposal, as discussed
above, we can define a surface gravity $\kappa_{\l}$ via $\l^a
\nabla_a \l^b \= \kappa_{\l} \l^b$.  Constancy of $\kappa_{\l}$ on
$\Delta$ follows from the same arguments that were used in
\cite{abf2}.  For completeness, let us briefly recall the structure of
that proof.  First, using conditions on derivatives of $l,n$
introduced in the Definition, one can express the self-dual part of
the Riemann curvature in terms of $\kappa_{\l}, \d\kappa_{\l}, \mu$
(and another field which is not relevant to this discussion).
Comparing this expression to the standard Newman-Penrose expansion of
the self-dual curvature tensor in terms of curvature scalars \cite{pr},
and using the fact that certain curvature scalars vanish on $\Delta$
(see (\ref{ricci}) and (\ref{weyl})), one can conclude 
\be \label{4.13}
\d\kappa_{\l}\wedge n \= 0, \quad{\rm and}\quad  
\kappa_{\l} \= \frac{\Psi_2}{\mu}.
\ee
The first equation implies that $\kappa_{\l}$ is spherically symmetric.  
Hence, it only remains to show that ${\cal L}_{\l}\,\kappa_{\l} \=0$.
Since $\mu$ is now a constant on $\Delta$, it suffices to show that
${\cal L}_\l \Psi_2  =0$. Now, the (second) Bianchi identity implies that
\be \label{bi}
{\cal L}_{\l}\, \left(\Psi_2 - \Phi_{11} \right) \= 0 \ee
Finally, using (\ref{3.2}) and the fact that the dilaton field $\phi$
is constant on $\Delta$, we conclude: $\Phi_{11} = - 8\pi^2 G\,(\exp
(2\alpha \phi_\Delta))\, (\Q_\Delta/a_\Delta)^2$.  
Thus, $\Phi_{11}$ is
constants on $\Delta$.  Combining these results, we conclude ${\cal
L}_{\l} \kappa_{\l} \= 0$, whence $\kappa_{\l}$ is constant on
$\Delta$.  This establishes the zeroth law.

With an eye toward the first law, let us now carry out the second step
in fixing the normalization of $\l$.  So far, we have only required
that $\mu$ be a (negative) constant but not fixed its value.  Under
the rescaling $\mu \mapsto c^{-1}\mu$ we have: $\l \mapsto c\l$, and
$\kappa_{\l} \mapsto c \kappa_{\l}$.  Hence, the remaining rescaling
freedom in $\l$ and $\kappa$ can be exhausted simply by fixing the
value of $\mu$ in terms of the isolated horizon parameters.  The
obvious strategy is to fix $\mu$ to the value $\mu_{\rm gGM}$ it takes
on the gGM solutions.  However, there is a practical difficulty:
Although $\mu_{\rm gGM}$ is a well-defined function of the isolated
horizon parameters $R_\Delta,{\Q}_\Delta$ and $\phi_\Delta$, to our
knowledge, there is no closed expression for it in terms of these
parameters.

Therefore, for computational convenience, let us first make a change
of coordinates on the three dimensional parameter space ${\cal P}$
defined by the triplet $(R_\Delta, \Q_\Delta, \phi_\Delta)$. Introduce
$r_\Delta$ and $k_\Delta$ via
\ba \label{new}
 R_\Delta &=&  r_\Delta\, \left[1 - (1+\alpha^2)\,
\frac{G\Q_\Delta^2 e^{2\alpha k_\Delta}}
{r_\Delta^2}\right]^{{\alpha^2}/{1+\alpha^2}}
\nonumber\\
e^{\alpha \phi_{\Delta}} &=& \frac{R_\Delta}{r_\Delta}\, 
e^{\alpha k_\Delta}.
\ea
It is easy to check that $(r_\Delta, \Q_\Delta, k_\Delta)$ is a good
set of coordinates on ${\cal P}$ with ranges $r_\Delta^2  >
{G}\Q_\Delta^2, \, -\infty <\Q_\Delta < \infty$ and $-\infty <
k_\Delta < \infty$.  With these definitions, in the gGM solutions we
have:
\be
r_\Delta = r_+,\quad \Q^2_\Delta =
\left[\frac{r_- r_+}{1+\alpha^2}\right]
e^{-2\alpha k}, \quad {\rm and} \quad k_\Delta =k\, .
\ee
However, from a conceptual point of view, on general isolated horizons
$(R_\Delta, \Q_\Delta, \phi_\Delta)$ are the fundamental parameters
and $(r_\Delta, \Q_\Delta, k_\Delta)$ are simply convenient functions
of them defined via (\ref{new}). In particular, on a general isolated
horizon, $k_\Delta$ is unrelated to the value $\phi_\infty$ of the dilaton
field at infinity.

We are now ready to express $\mu_{\rm gGM}$ of (\ref{kappa1}) in a closed 
form:
\be \label{mu}
\mu_{\rm gGM} \= -\frac{1}{r_\Delta}\left[1 - 
\frac{G\Q^2e^{2\alpha k_\Delta}}{r_\Delta^2}\right] 
\left[1 - (1+\alpha^2) \,
\frac{G\Q^2e^{2\alpha k_\Delta}}{r_\Delta^2}\right]^{-1} 
\ee
Therefore, for \textit{general isolated horizons} we will gauge-fix
$\l$ by demanding that $\mu$ be given by the right side of (\ref{mu}).
With this final `gauge-fixing', $\l$ is uniquely determined on each
isolated horizon. We will denote the corresponding $\kappa_{\l}$
simply by $\kappa$ and call it \textit{the surface gravity} of the
isolated horizon $\Delta$. Using (\ref{4.13}), (\ref{ricci}) and
(\ref{weyl}) and the expression (\ref{mu}) of $\mu$, we can express
$\kappa$ in terms of the isolated horizon parameters as
\be \label{kappa}
\kappa = \frac{2\pi}{a_\Delta}\, r_\Delta\, \left[1- (1+\alpha^2)\, 
\frac{G\Q_\Delta^2\,e^{2\alpha k_\Delta}}{r_\Delta^2}\right]\, ,
\ee
where, as before, $a_\Delta = 4\pi R_\Delta^2$ is the area of the
horizon. By construction, in the gGM solutions, $\kappa$ reduces to
(\ref{kappa1}).

Finally, as discussed in \cite{abf2}, to arrive at the standard form
of the first law, we need to gauge fix the electro-magnetic potential
$\Phi\= \bA\cdot \l$.  Again, the strategy is similar to the one we
used to gauge fix the normalization of $\l$: We first express
$\Phi_{\rm gGM}$ in the gGM solutions in terms of the isolated horizon
parameters and then ask that $\Phi$ be the same function of the
parameters in the general case.  Again, to obtain an expression of
$\Phi_{\rm gGM}$ in a closed form, we are led to use $(r_\Delta,
\Q_\Delta, k_\Delta)$.  It is easy to verify that
\be  \label{Phi}
\Phi_{\rm gGM} = \frac{\Q_\Delta}{r_\Delta}\, e^{2\alpha k_\Delta}\, .
\ee
Hence, on general isolated horizons, we work in a gauge in which
$\Phi$ is given by the right side of (\ref{Phi}).  As in the
Einstein-Maxwell case \cite{abf2}, this gauge fixing makes the
matter action $S_{\rm Dil}$ functionally differentiable even in
presence of the internal boundary $\Delta$.

We conclude this section with a remark.  A general isolated horizon
carries three independent parameters, $(R_\Delta, \Q_\Delta,
\phi_\Delta)$, arising from the metric, Maxwell and dilaton fields
respectively. Therefore, to unambiguously implement the above
procedure to gauge-fix the normalization of $\l$, we needed access to
a three parameter family of `reference' static solutions, one for each
point in the parameter set. This is why it was essential to consider
the generalization of the Gibbons-Maeda solutions to include non-zero
values of $\phi_\infty$. Had we restricted ourselves to the original
Gibbons-Maeda family, the mapping between the 3-dimensional parameter
space ${\cal P}$ and the space of static solutions would have been
ambiguous. For the generalized Gibbons-Maeda family, the mapping is
1-1 and hence unambiguous.

\section{Mass and the first law}
\label{s5}

As explained in the Introduction, since space-times under
consideration are not necessarily stationary, it is no longer
meaningful to identify the ADM mass $M$ with the mass $M_\Delta$ of
the isolated horizon.  For the formulation of the first law,
therefore, we must first introduce an appropriate definition of
$M_\Delta$.  The Hamiltonian framework provides a natural strategy.
In the Einstein-Maxwell case the total Hamiltonian consists of a bulk
term and \textit{two} surface terms, one at infinity and the other at
the isolated horizon.  As usual, the bulk term is a linear combination
of constraints and the surface term at infinity yields the ADM energy.
In a rest-frame adapted to the horizon it is then natural to identify
the surface term at $\Delta$ as the horizon mass, $M_\Delta$.  Indeed,
there are several considerations that support this identification
\cite{abf2}. We will use the same strategy in the dilatonic case.

\begin{figure}
\centerline{
\hbox{\psfig{figure=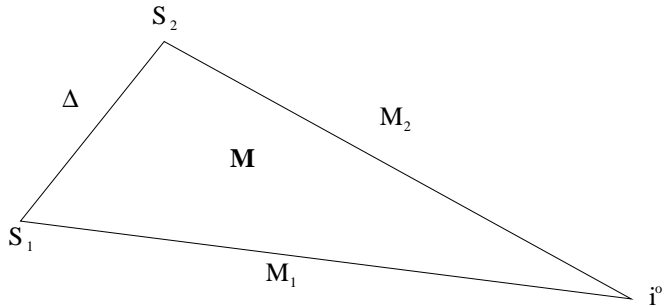,height=4cm}}}
\bigskip
\caption{Region $\M$ of space-time considered in the
variational principle is bounded by two partial Cauchy surfaces $M_1$
and $M_2$.  They intersect the isolated horizon $\Delta$ in preferred
2-spheres $S_1$ and $S_2$ and extend to spatial infinity $\spi$.}\label{exam}
\end{figure}

For the gravitational part of the action and Hamiltonian, the
discussion of \cite{abf2} only assumed that the stress-energy tensor
satisfies two conditions at $\Delta$: i) $-T_{ab}\l^a$ is a future
pointing causal vector field on $\Delta$; and, ii) $T_{ab}\l^a n^b$ is
spherically symmetric on $\Delta$.  Both these conditions are met in
the present case.  Therefore, we can take over the results of
\cite{abf2} directly.  For the matter part of the action and
Hamiltonian, the overall situation is again analogous, although there
are the obvious differences in the detailed expressions because of
dilaton couplings. As in the Einstein-Maxwell case, matter terms
contribute to the surface terms in the Hamiltonian only because one
has to perform one integration by parts to obtain the Gauss constraint
in the bulk term.

The net result is the following.  Consider a foliation of the given
space-time region $\M$ by a 1-parameter family of (partial)
Cauchy surfaces $M$, each of which extends from the isolated horizon
$\Delta$ to spatial infinity $i^o$ (see Figure).  We will assume that
$M$ intersects $\Delta$ in a 2-sphere belonging to our preferred
foliation and that the initial data induced on $M$ are asymptotically
flat. Denote by $S_\Delta$ and $S_\infty$ the 2-sphere boundary of $M$
at the horizon and infinity, respectively.  Choose a time-like vector
field $t^a$ in $\M$ which tends to the unit time-translation
orthogonal to the foliation at spatial infinity and to the vector
field $\l^a$ on $\Delta$, normalized as in Section \ref{s4}.  Then,
the Hamiltonian $H_t$ generating evolution along $t^a$ is given by:
\ba \label{ham}
H_t &=& \int_M {\rm constraints}\,\,+\,\, \lim_{R_o\to \infty}
\oint_{S_{R_o}}\left(\frac{R_o}{4\pi G}\, {\Psi_2}\right) \,\,
{}^2\epsilon  \nonumber\\
&+& \oint_{S_\Delta} \left(\frac{\mu^{-1}}{4\pi G} \Psi_2 \right) 
{}^2\epsilon + (\bA\cdot l){\Q}_\Delta\nonumber\\
\ea
where $S_{R_o}$ are large 2-spheres of radius $R_o$. (The calculation
and the final result are completely analogous to those in the
Einstein-Maxwell case \cite{abf2}. The only differences are: i) Since
we have set the cosmological constant equal to zero, the scalar
curvature of the 4-metric vanishes on $\Delta$, and ii) the surface
term that results from integration by parts of the Maxwell Gauss
constraint now contains $\Q_\Delta$ in place of ${Q}_\Delta$.)  Note
that the surface terms depend only on the `Coulombic' parts of the
gravitational and Maxwell fields.  The form of their integrands is
rather similar since asymptotically the Newman-Penrose
spin-coefficient $\mu$ goes as $R_o^{-1}$.  However, while the surface
term at infinity depends \textit{only} of the Weyl curvature, the term
at the horizon depends also on the Maxwell Potential.

It is easy to check that the surface term at infinity is, as usual,
the time component $P_a^{\rm ADM}t^a$ of the ADM 4-momentum $P_a^{\rm
ADM}$, which in the present -,+,+,+ signature is negative of the ADM
energy, $P_a^{\rm ADM} t^a = -E^{\rm ADM}$.  It is natural to identify
the surface term at $S_\Delta$ as the energy of the isolated horizon.
(There is no minus sign because $S_\Delta$ is the \textit{inner}
boundary of $M$).  Since $t^a \= \l^a$ and since $\l^a$ represents the
`rest frame' of the isolated horizon, this energy can in turn be
identified with the horizon mass $M_\Delta$.  Thus, we have:
\be \label{mass}
 M_\Delta = \oint_{S_\Delta} \left(\frac{\mu^{-1}}{4\pi G} \Psi_2 \right) 
{}^2\epsilon + (\bA\cdot l){\Q}_\Delta \, .
\ee

Using the expression (\ref{4.13}) of surface gravity in terms of the
Weyl tensor and the expression (\ref{Phi}) of $\bA\cdot \l$ on
$\Delta$, we can cast $M_\Delta$ in a more familiar form:
\be \label{smarr}
M_\Delta = \frac{1}{4\pi G}\, \kappa a_\Delta\, +\, \Phi \Q_\Delta 
\ee
Thus, as in the gGM solutions, we obtain the Smarr formula.  However,
the meaning of various symbols in the equation is somewhat different.
Since an isolated horizon need not be a Killing horizon, in general
$M_\Delta$ does \text{not} equal the ADM mass, nor is $\kappa$ or
$\Phi$ computed using a Killing field.  If the space-time happens to
be static with $t^a$ as the Killing field, general arguments from
symplectic geometry can be used to show that the numerical value of
the Hamiltonian $H_t$ vanishes at this solution \cite{abf2}.  Since
the constraints are satisfied in any solution, the bulk term in
(\ref{ham}) vanishes as well.  Hence, in this case, $M_\Delta = E^{\rm
ADM}$.  In general, however, the two differ by the `radiative energy'
in the space-time. In the Einstein-Maxwell case, one can argue
\cite{abf2} that, if $\Delta$ stretches all the way to $i^+$, 
under suitable regularity conditions $M_\Delta$ equals the future
limit of the Bondi-energy in the frame in which the black hole is at
rest. There appears to be no obstruction to extending that argument to
the dilatonic case. Finally, as emphasized in \cite{abf2}, the matter
contribution to the mass formula (\ref{mass}) is subtle: while it does
not include the energy in radiation outside the horizon, it does
include the energy in the `Coulombic part' of the field associated
with the black hole hair. (Recall that the future limit of the Bondi
energy has this property.) However, since this issue was discussed in
detail in \cite{abf2}, and since dilaton couplings do not add
substantive complications, we will not repeat that discussion it here.

We now have expressions of the mass $M_\Delta$, surface gravity
$\kappa$, area $a_\Delta$ and the electric potential $\Phi$ of any
isolated horizon in terms of the convenient functions $r_\Delta,
\Q_\Delta, k_\Delta$ of its fundamental parameters $R_\Delta,
\Q_\Delta, \phi_\Delta$:
\ba \label{functions1}
M_\Delta=\frac{r_\Delta}{2G}\left(1+ (1-\alpha^2)
\frac{G{\Q}_\Delta^2\, e^{2\alpha k_\Delta}}{r^2_\Delta}\right)
\quad &;&\quad 
a_\Delta=4\pi r_\Delta^2\left(
1- (1+\alpha^2)\, \frac{G\Q^2_\Delta\, e^{2\alpha k_\Delta}}
{r_\Delta^2}\right)^{{2\alpha^2}/{(1+\alpha^2)}}
\nonumber\\
\kappa=\frac{2\pi r_\Delta}{a_\Delta}\, \left(1-
(1+\alpha^2)\frac{G{\Q}_\Delta^2\, e^{2\alpha k_\Delta}}{r_\Delta^2}
\right)\quad &;&\quad
\Phi=\frac{{\Q}_\Delta}{r_\Delta}\, e^{2\alpha k_\Delta}\, .
\ea
Since all quantities are now expressed in a closed form in terms of
the same parameters of isolated horizons, it is easy to compute their
variations. A simple calculation yields%
\footnote{A similar expression of the first law with contributions to
the `work term' from scalar fields was first obtained in \cite{gkk}.
That analysis was restricted only to \textit{static solutions} but
carried out on a more general context of theories with
\textit{several} scalar fields.}
\ba  \label{1law2} 
\delta{M_\Delta} &=& \frac{1}{8\pi G}\, \kappa \delta{a_\Delta} + \Phi 
\delta{\Q_\Delta} + \alpha \Phi \Q_\Delta\, \delta k_\Delta
\nonumber\\
&=&\frac{1}{8\pi G}\, \kappa \delta{a_\Delta} + \hat\Phi 
\delta{\hat{Q}_\Delta}
\ea
where $\hat{\Phi} = \Phi e^{-\alpha k_\Delta}$ (i.e., ${\hat{\Phi}}^2 =
\Q Q[S_\Delta]/R^2_\Delta$) and $\hat{Q} = {\Q}e^{\alpha
k_\Delta}$. Thus, as one might have expected from the Einstein-Maxwell
analysis of \cite{abf2}, the first law (\ref{gGM1law}) for general
isolated horizons has the same form as (\ref{gGM1law}), the first law
in gGM solutions. The only difference in the detailed expression is
the replacement of the ADM mass $M$ by the horizon mass $M_\Delta$.

However, physically, this difference is important \cite{abf2}.
Consider, for example, a space-time which admits an isolated horizon
$\Delta_1$ for a finite time interval which ceases to be isolated for
a small duration because of the influx of radiation into the horizon
and settles down again to an isolated horizon $\Delta_2$. Then, using
two `triangular' regions ${\bf M_1}$ and ${\bf M_2}$ in the space-time
with internal boundaries $\Delta_1$ and $\Delta_2$ respectively (see
Fig. 1), one can compute the masses $M_{\Delta_1}$ and $M_{\Delta_2}$
of the two isolated horizons. While the ADM mass $M$ of the space-time
remains unchanged during this physical process, the isolated horizon
mass does change; $M_{\Delta_1} \not= M_{\Delta_2}$. By using the
Raychaudhuri equation, one can show that the the difference is
precisely the mass associated with the radiation which fell in. Thus,
in this process the mass contained in the radiation outside the
horizon decreases, the horizon mass increases but the sum, which
equals the ADM mass $M$, remains unchanged. If there is flux of
charged matter across the horizon, the situation is more subtle but
the final result is the same. As discussed in detail in \cite{abf2},
in this case, there does not appear to exist a treatment of the
physical process version of the first law without recourse to the
isolated horizon framework.

Finally, note that the expression (\ref{smarr}) of $M_\Delta$ can be
trivially rewritten in terms of the hatted variables:
\be
 M_\Delta = \frac{1}{4\pi G}\, \kappa a_\Delta\, +\, \hat\Phi 
\hat{Q}_\Delta 
\ee
Thus, for dilatonic isolated horizons, the smarr formula and the first
law have the same form as in the Einstein-Maxwell case, the only
difference being that the electric potential and charge are now
replaced by $\hat\Phi$ and $\hat{Q}$. 

We will conclude this discussion with three remarks:
\hfill\break
a) The space ${\cal IH}$ of space-times admitting isolated horizons is
infinite dimensional. Fortunately, however, the boundary conditions
are sufficiently strong to enable us to express physical parameters
such as the mass $M_\Delta$ and the surface gravity $\kappa$ in terms
of only three parameters; they are lifts to ${\cal IH}$ of functions
on ${\cal P}$.  Explicit calculations presented in this section made a
heavy use of the new coordinates $(r_\Delta,\Q_\Delta, k_\Delta)$ on
${\cal P}$. However, from a conceptual standpoint, the introduction of
these coordinates is not essential. For, $M_\Delta$, $\kappa$,
$a_\Delta$ and $\Phi$ are all well-defined functions on ${\cal P}$ and
the first law is a statement of relations between the gradients of
three of these functions. It makes no reference at all to the choice
of coordinates on ${\cal P}$. Indeed, in principle, we could have used
only the fundamental parameters $(R_\Delta, \Q_\Delta, \phi_\Delta)$
throughout; passage to $(r_\Delta, \Q_\Delta, k_\Delta)$ served to
simplify the algebra.

b) The new parameterization entered our explicit calculations because
we wished to reduce the rescaling freedom in $\l$ in such a way that
in the static (gGM) solutions $\l$ coincides with the restriction to
$\Delta$ of that static Killing field which is unit at infinity. To
implement this condition, we had to ask that, on general isolated
horizons, $\mu$ be the same function of parameters as $\mu_{\rm gGM}$
is in static solutions and $\mu_{\rm gGM}$ could not be expressed in a
closed form in terms of $(R_\Delta, \Q_\Delta, \phi_\Delta)$. While
this choice of scale for $\l$ is most convenient for making contact
with the standard static framework, we could have made other
choices. In general, if we rescale $\l$ via $\l \mapsto \l^\prime = c
\l$, where $c$ is a constant on $\Delta$ (but may be a function of the
parameters), we have $\mu \mapsto \mu^\prime = c^{-1}\mu$, $M_\Delta
\mapsto M^\prime_\Delta = c M_\Delta$, $\kappa \mapsto \kappa^\prime =
c \kappa$ and $\Phi \mapsto \Phi^\prime = c\Phi$ while the area
$a_\Delta$ and the charge $\Q_\Delta$ remain unchanged (see Eq
(\ref{mass}) and the discussion in Section \ref{s4}). 

Let us choose $c = R_\Delta/r_\Delta$. Then we have
\ba \label{functions2}
M_\Delta^\prime =\frac{R_\Delta}{2G}\left(1+ (1-\alpha^2)
\frac{G{Q^\prime}_\Delta^2\,}{R^2_\Delta}\right)
\quad &;&\quad 
a_\Delta^\prime =4\pi R_\Delta^2
\nonumber\\
\kappa^\prime =\frac{2\pi R_\Delta}{a_\Delta}\,\, \left(1-
(1+\alpha^2)\frac{G{Q^\prime}_\Delta^2\,}{R_\Delta^2}
\right)\quad &;&\quad
\Phi^\prime = \frac{Q[S_\Delta]}{R_\Delta}\, ,
\ea
where, as before, $Q[S_\Delta] = \Q_\Delta\, e^{2\alpha\phi_\Delta}$
and ${Q^\prime}^2_\Delta = \Q_\Delta Q[S_\Delta]$.  Thus, in this
rescaling gauge, all physical quantities are expressed entirely in
terms of the fundamental parameters $(R_\Delta, \Q_\Delta,
\phi_\Delta)$. In this gauge the first law becomes:
\be \label{1law3}
\delta{M^\prime_\Delta} = \frac{1}{8\pi G}\kappa^\prime\, \delta
a_\Delta^\prime + (1-\alpha^2) \hat{\Phi^\prime} \delta 
Q^\prime_\Delta\, ,
\ee 
where $\hat{\Phi^\prime} = Q^\prime_\Delta/R_{\Delta}$. Note that Eqs
(\ref{functions2}) and (\ref{1law3}) are completely equivalent to Eqs
(\ref{functions1}) and (\ref{1law2}); the two sets express the same
information but in different rescaling-gauges. One set is adapted to
the fundamental parameters of isolated horizons but fails to reduce to
the standard set of expressions used in the static context.  The other
set does reproduce the standard expressions in the static case but
involves the use of auxiliary parameters $(r_\Delta, \Q_\Delta,
k_\Delta)$.

c) Our main results can be immediately generalized to incorporate a
cosmological constant $\Lambda$.  The only essential change in Section
\ref{s3} is that the scalar curvature $R$ of the 4-metric at the
horizon is now given by $R=4\Lambda$, which in turn introduces factors
of $\Lambda$ in the expressions of physical quantities in Sections
\ref{s4} and \ref{s5} (see \cite{abf2}).  Now, it is known that there
are no static black hole solutions in presence of a positive
cosmological constant \cite{constant}. Therefore, if isolated horizons
exist with $\Lambda >0$, we would not be able to fix the normalization
of $\l$ using the procedure used in Section \ref{s4}. However, as the
discussion of Sections \ref{s4} and \ref{s5} and the discussion in the
point b) above shows, one could still establish the zeroth and first
laws. For $\Lambda <0$, static solutions are known to exist
\cite{constant} and we can repeat the procedure given above. Then, one
can show
\be 
\kappa= \frac{2\pi r_\Delta}{a_\Delta}\left(1-(1+\alpha^2)
\frac{G\tilde{Q}^2}{r_\Delta^2}e^{2\alpha k_\Delta}-\frac{a_\Delta}{4\pi}
\Lambda\right)
\ee
The electrostatic potential $\Phi$ is unaffected and the Smarr formula
(\ref{smarr}) and the first law continues to hold.

\section{Discussion}
\label{s6}

In \cite{abf2}, the zeroth and first laws were established for
mechanics of non-rotating isolated horizons in general relativity.
While the restrictions on the matter content were rather mild, that
analysis did not allow a dilatonic charge or dilaton couplings.  In
this note, we extended that analysis to dilaton gravity.  Key ideas
and strategies are the same as those introduced in \cite{abf2}.
However, the incorporation of general dilaton couplings ---i.e.,
arbitrary values of the coupling parameter $\alpha$--- was not
entirely straightforward because the specific gauge choices needed to
define surface gravity and electro-magnetic potential on general
isolated horizons are now rather complicated.  However, once an
appropriate viewpoint towards parameterization of general isolated
horizons is introduced, one realizes that the complications are rather
tame; only the algebra becomes more involved.  In the final picture,
the proof of the zeroth law is completely analogous to that in the
Einstein-Maxwell case and the first law has exactly the same form as
in static gGM solutions.

There are three natural directions in which the results of this paper
could be extended.  First, as in the previous work \cite{ack,abf2}, our
boundary conditions imply that the intrinsic 2-metric on $\Delta$ is
spherically symmetric (although the space-time metric need not admit
\textit{any} Killing field in a neighborhood of $\Delta$.)  It is of
interest to extend the analysis to incorporate (non-rotating but)
distorted isolated horizons.  This would only require weakening of the
boundary condition (iii.b) on derivatives of $n^a$ and (iv) on
spherical symmetry of $\phi_1$.  A second and more interesting
extension would be inclusion of rotation.  Again, only conditions
(iii.B) and (iv) would have to be weakened. Very recently, both these
extensions were carried out for the Einstein-Maxwell theory
\cite{afk,abl}. Furthermore, in contrast to the present paper, these
analyses do not fix the scaling of $\l^a$ using static solutions; a
more general strategy is involved. Therefore, it is likely that those
considerations will extend to dilaton couplings as well, even though
explicit distorted or rotating dilatonic solutions are still not
known. Finally, as pointed out at the end of Section \ref{s3},
isolated horizons provide a natural home to study black holes with
non-standard asymptotic structure. For theories considered in
\cite{chm}, the present analysis already suffices to establish the
zeroth law, i.e., the constancy of $\kappa$ on the horizon. It would
be interesting to analyze whether the methods developed in
\cite{afk,abl} can be used to establish the first law.

\section*{acknowledgments}

We would like to thank Chris Beetle, Steve Fairhurst and Jerzy
Lewandowski for discussions, Hernando Quevedo for checking certain
properties of the Gibbons-Maeda solutions using REDUCE, and Gary
Gibbons and a referee for pointing out references \cite{gkk} and
\cite{chm} respectively. This work was supported in part by the NSF
grants PHY94-07194, PHY95-14240, INT97-22514 and by the Eberly
research funds of Penn State.  AC was also supported by DGAPA-UNAM
grant IN121298 and by CONACyT Proy. Ref. I25655-E.

\end{document}